\documentclass[epj]{svjour}

\usepackage{graphics}
\usepackage{psfrag}
\usepackage{epsfig}
\usepackage{epsf}
\usepackage{float}
\usepackage{amssymb,stmaryrd,latexsym}
\usepackage[utf8x]{inputenc}
\usepackage{amsmath,amssymb,mathrsfs}
\usepackage{graphicx}
\usepackage{slashed}
\usepackage{bbm}
\usepackage{fancyhdr}
\usepackage{a4wide}
\usepackage{dsfont}
\usepackage[usenames]{color}
\usepackage{lscape}
\newcommand{\non}{\nonumber}

\title{\hfill {\tiny FZJ-IKP-TH-2010-20, HISKP-TH-10/24}\\
Light meson mass dependence of the positive parity heavy-strange mesons}

\author{Martin~Cleven$^1$
      , Feng-Kun Guo$^1$
      , Christoph~Hanhart$^{1,2}$
      , and Ulf-G. Mei{\ss}ner$^{1,2,3}$ 
\\
{\small $\rm ^1$Institut f\"{u}r Kernphysik and J\"ulich Center
             for Hadron Physics, Forschungszentrum J\"{u}lich,
             D--52425 J\"{u}lich, Germany
\\
$\rm ^2$Institute for Advanced Simulation,
             Forschungszentrum J\"{u}lich, D--52425 J\"{u}lich, Germany
\\
$\rm ^3$Helmholtz-Institut f\"ur Strahlen- und
             Kernphysik and Bethe Center for Theoretical Physics, Universit\"at
             Bonn,  D--53115 Bonn, Germany}}

\authorrunning{Cleven et al.}

\titlerunning{Light meson mass dependence of the positive parity heavy-strange mesons}


\begin{document}

\abstract{ We calculate the masses of the resonances $D_{s0}^*(2317)$ and
$D_{s1}(2460)$ as well as their bottom partners as bound states of a kaon and a
$D^{(*)}$- and $B^{(*)}$-meson, respectively, in unitarized chiral perturbation
theory at next-to-leading order. After fixing the parameters in the
$D_{s0}^*(2317)$ channel, the calculated mass for the $D_{s1}(2460)$ is found in
excellent agreement with experiment. The masses for the analogous states with a
bottom quark are predicted to be   $M_{B^*_{s0}}=(5696\pm 40)~{\rm MeV}$ and
$M_{B_{s1}}=(5742\pm 40)~{\rm MeV}$ in reasonable agreement with previous
analyses.
In particular, we predict $M_{B_{s1}}{-}M_{B_{s0}^*}=46\pm 1 \ \mbox{MeV}
.$
 We also explore the
dependence of the states on the pion and kaon masses. We argue that the kaon mass dependence of
a kaonic bound state should be almost {\em linear with slope about unity}. Such
a dependence is specific to the assumed molecular nature of the states. We
suggest to extract the kaon mass dependence of these states from lattice QCD calculations.}

\maketitle

\section{Introduction}

The discovery of two narrow resonances with open charm in 2003 opened a new
chapter in hadronic spectroscopy. The $D_{s0}^*(2317)$ was discovered by the
BaBar collaboration \cite{Aubert:2003fg} and shortly after the $D_{s1}(2460)$ by the
CLEO collaboration~\cite{Besson:2003cp}.\footnote{BaBar  already observed a
  signal of the $D_{s1}(2460)$ in~\cite{Aubert:2003fg}.} Both
were confirmed later.

These resonances were not in agreement with the predictions of
constituent quark models, for instance the Godfrey-Isgur
model~\cite{Godfrey:1985xj}, 
that on the other hand were very successful for all other open charm states
discovered so far.  The  masses of the $D_{s0}^*(2317)$ and of the $D_{s1}(2460)$
are approximately 160~MeV and 70~MeV, respectively, below the predictions in
Ref.~\cite{Godfrey:1985xj}.  In addition,
\begin{eqnarray}
M_{D_{s1}(2460)}{-}M_{D_{s0}^*(2317)}\simeq M_{D^*}{-}M_D \ .
\end{eqnarray}
This equality is difficult to explain in a conventional quark model. It can be
explained in a parity doubling model~\cite{Bardeen:2003kt,Nowak:2003ra}. This
result is preserved at one-loop level~\cite{Mehen:2005hc}. However, it is not
clear if the parity doubling is a consequence of quantum chromodynamics (QCD).
The equality of mass splittings appears naturally, if both states are assumed
to be molecular states of a kaon and a $D$ or a $D^*$~\cite{Guo:2009id}, since
spin-dependent interactions for a heavy quark are suppressed by two powers of
$\Lambda_{\rm QCD}/m_Q$ (see discussion in Section~\ref{sec:amplitudes}), with
$\Lambda_{\rm QCD}$ the soft QCD scale and $m_Q$ the mass of the heavy quark. Soon
after the discovery, based on phenomenological calculations, Barnes and
Close~\cite{Barnes:2003dj} and van Beveren and Rupp~\cite{vanBeveren:2003kd}
pointed out the possible molecular nature of these resonances: they described
them as bound states of $D^{(*)}K$ mesons.  For reviews of possible
explanations of these states, see Refs.~\cite{Swanson:2006st,Zhu:2007wz}. The
pure\-ly phenomenological calculations were shortly after improved by
Refs.~\cite{Kolomeitsev:2003ac,Hofmann:2003je,Guo:2006fu,Guo:2006rp}, where
variants of unitarized chiral perturbation theory were used, and in addition
open bottom molecules were also studied. In
Refs.~\cite{Gamermann:2006nm,Gamermann:2007fi} both resonances emerged as
molecules from Weinberg-Tomozawa type interactions using flavor-SU(4)
constraints.

Clearly it is important to find observables that allow one to identify the
nature of the mentioned states model-independently either from data or from a
comparison to lattice QCD results. Up to now a few such quantities were
pointed out, namely the value of the hadronic width of the
resonances~\cite{Faessler:2007gv,Lutz:2007sk,Guo:2008gp} ---  which in a
molecular assignment turns out to be significantly larger than that for a quark
state~\cite{Colangelo:2003vg} --- and the value of the $KD$ scattering length in
the resonance channel~\cite{Guo:2009ct} --- which gets large and negative, if
the states are molecules~\cite{Weinberg:1965zz,Baru:2003qq}. In this work we add an additional quantity, namely the
quark mass and especially the kaon mass dependence of the resonance masses. We
demonstrate that this dependence is qua\-litatively different for quark states and
for hadronic molecules and thus, a study of this dependence using lattice QCD is
of high importance.

Our study is based on Refs.~\cite{Guo:2008gp,Guo:2009ct}, but extended
in that now $D$- and $D^*$-mesons are both included on equal footing. This allows
us to also investigate the implications of heavy quark spin symmetry.

The paper is organized as follows. In Section~\ref{sec:amplitudes} we give the
Lagrangian for and display the calculation of Goldstone boson scattering off
$D$- and $D^*$-mesons as well as the unitarization scheme used to generate
the $D_{s0}^*(2317)$ and the $D_{s1}(2460)$ as hadronic molecules.  In
Section~\ref{sec:chextra} the coupled channel problem in the strange\-ness $S=1$
isoscalar channel is solved and the $D_{s0}^*(2317)$ and the $D_{s1}(2460)$
are generated as $DK$ and $D^* K$ molecules. We discuss the extra\-polation of
their masses to unphysical values for the pion and kaon mass, respectively. A
brief summary and outlook is given in Section~\ref{sec:summary}. The appendix
contains some details on the derivation of the pertinent scattering amplitudes.

\section{Scattering amplitudes}\label{sec:amplitudes}

The leading order (LO) Lagrangian in Heavy Meson Chiral Perturbation Theory is
given by the kinetic energy of the heavy fields, the coupling of the heavy
fields to pions and the mass splitting of the heavy
mesons~\cite{Burdman:1992gh,Wise:1992hn,Yan:1992gz}
\begin{eqnarray}\nonumber
\mathscr L_{\rm LO}&=&-i {\rm Tr}[\bar H_a v_\mu D^\mu_{ba} H_b]
+ g_\pi {\rm Tr}[\bar H_aH_b\gamma_\nu\gamma_5] u^\nu_{ba}\\
&&+\frac{\lambda}{m_Q} {\rm Tr}[\bar H_a\sigma_{\mu\nu}H_a\sigma^{\mu\nu}]
\end{eqnarray}
with the heavy fields defined as \cite{Falk:1990yz}
\begin{eqnarray}\nonumber
 H=\frac{1+\slashed{v}}{2}\left[\slashed{V}+iP\gamma_5\right], \quad \bar{\,H}=\gamma^0H^\dagger\gamma^0,
\end{eqnarray}
where Latin indices denote flavor and the trace is taken over the Gamma matrices.
Here, the heavy fields are $P=(D^0,D^+,D_s^+)$ and
$V_\mu=(D^{*0}_{\mu},D^{*+}_{\mu},D_{s,\mu}^{*+})$, and
$v_\mu=(1,\vec{0})+\mathscr O(\vec p/M)$ is the heavy mesons ve\-lo\-ci\-ty. The
covariant derivative is
\begin{eqnarray}
D_\mu=\partial_\mu+\Gamma_\mu \, , \quad
\Gamma_\mu=\frac{1}{2}\left(u^\dagger\partial_\mu u+u\partial_\mu u^\dagger\right)
\end{eqnarray}
where
\begin{eqnarray}
 U=\exp\left(\frac{\sqrt{2}i\phi}{F}\right), \quad u^2=U
\label{Udef}
\end{eqnarray}
with $F$ the pion decay constant in the chiral limit. Up to the order we are
working, $F$ can be replaced by the physical pion decay constant
$F_\pi=92.4~\rm{MeV}$. As we will see, it is basically $F$ that
provides the strength of the interaction.

The Goldstone boson fields are collected in the matrix
$\phi$ with
\begin{eqnarray} 
&\phi= \left( \begin{array}{ccc}
\frac{1}{\sqrt{2}}\pi^0+\frac{1}{\sqrt{6}}\eta & \pi^+ & K^+ \\
\pi^- & -\frac{1}{\sqrt{2}}\pi^0+\frac{1}{\sqrt{6}}\eta &  K^0 \\
K^- & \bar{K^0} & -\frac{2}{\sqrt{6}}\eta \end{array} \right).
\end{eqnarray}
The axial-vector current is defined as
\begin{eqnarray}
u_\mu=iu^\dagger(\partial_\mu U)u^\dagger.
\end{eqnarray}
The low-energy constant (LEC) $g_\pi$ can be fixed from the strong decay of the
$D^{*+}$. Taking the values from the Particle Data Group (PDG) \cite{pdg2010}, we
find $g_\pi=0.30\pm0.03$.

The spin-symmetry breaking term gives rise to a mass difference
\begin{eqnarray}
\label{eq:deltadef}
 \Delta=M_{V}-M_P=-8\frac{\lambda}{m_Q}~,
\end{eqnarray}
with $m_Q$ the pertinent heavy quark mass.
In this work we use physical values for the meson masses and thus
the mentioned term is considered automatically. Note, since
$M_{D_s*}-M_{D_s}\neq M_{D*}-M_{D}$ (c.f. Eqs.~(\ref{charmmasses})),
in this way we also include an effect of simultaneous spin symmetry
and SU(3) violation, which is formally of  next--to--next--to leading order (N$^2$LO). We come back
to the quantitative role of this subleading effect below.

\begin{figure*}[ht]
\centering
\includegraphics[width=0.3\linewidth]{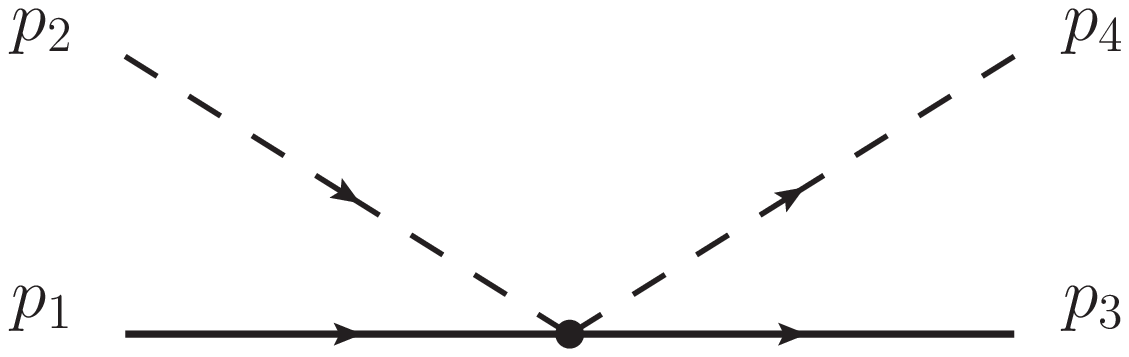} \hspace{0.5cm}
\includegraphics[width=0.3\linewidth]{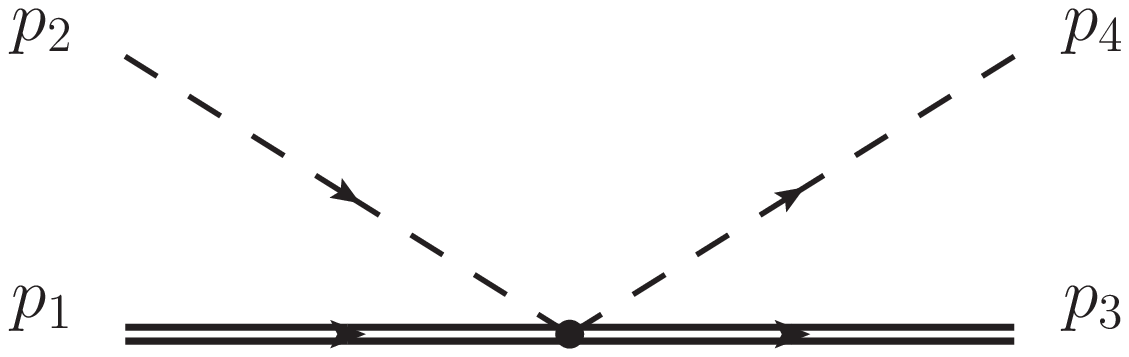}\hspace{0.5cm}
\includegraphics[width=0.3\linewidth]{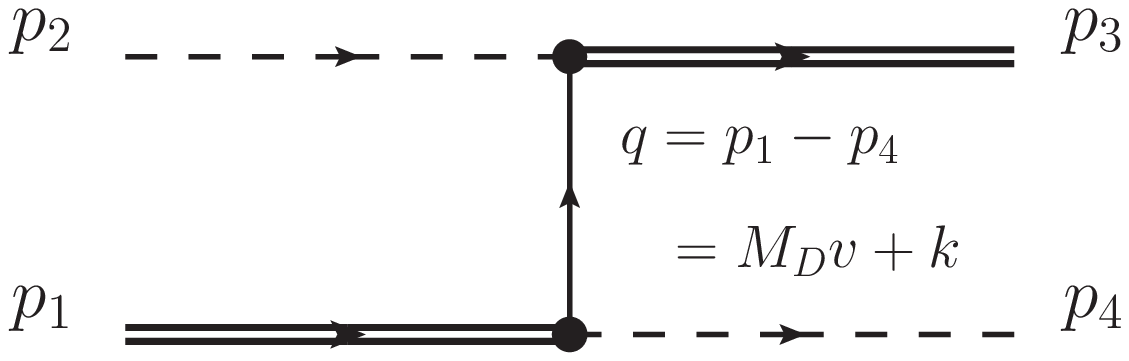}
\caption{Diagrams contributing at LO and NLO to Goldstone boson scattering off 
$D$- and $D^\star$-mesons and the pertinent kinematics. Dashed lines denote Goldstone
bosons, solid lines charmed pseudoscalar mesons and solid double lines charmed
vector mesons, in order.} \label{fig:diagrams}
\end{figure*}

The Lagrangian at next-to-leading order (NLO) is
\begin{eqnarray}\nonumber\label{Eq:LagrNLO}
  \mathscr{L}_{\rm NLO}
&=&+ {\rm Tr}[\bar H_a(-h_0(\chi_+)_{bb}+h_2 (u_\mu u^\mu)_{bb}\\ \nonumber
&&+h_4(-iv_\mu)(u^\mu u^\nu)_{bb} (iv_\nu))H_a]\\ \nonumber
&&+ {\rm Tr}[\bar H_a (h_1(\tilde\chi_+)_{ba}+h_3 (u_\mu u^\mu)_{ba})H_b]\\
&&+h_5 {\rm Tr}[\bar H_a (-iv_\mu)\{u^\mu, u^\nu\}_{ba} (iv_\nu)H_b], 
\end{eqnarray}
where we have introduced explicit chiral symmetry breaking terms due to non-vanishing
light quark masses using
\begin{eqnarray}\nonumber
&\chi_\pm=u^\dagger\chi u^\dagger\pm u\chi^\dagger u ~,\\
&\tilde \chi_\pm=\chi_+-\frac{1}{3}(\chi_+)_{aa}~,
\end{eqnarray}
with $\chi\!=\!2B \cdot $diag$(m_u,m_d,m_s)$, $B=|\!\left<0|\bar
qq|0\right>\!|/F^2$ and $(\chi_+)_{aa}$ is the trace of $\chi_+$.

Following Ref.~\cite{Lutz:2007sk} we neglect the terms $h_0$, $h_2$ and
$h_4$ since they are suppressed in the large-$N_C$ limit because of the second
flavor trace. The LEC $h_1$ can be determined using the relation $B(m_s-\hat
m)=(M^2_{K^0}+M^2_{K^+})/2-M^2_{\pi^0}$ as done in \cite{Hofmann:2003je}. We use
the value $h_1=0.42\pm0.00$.

Since we are interested in the masses of the resonances, isospin breaking can be
neglected. Hence, we take averaged values for charged and neutral particle
masses, and the pion decay constant from the PDG, as listed in the
following~\cite{pdg2010}:
\begin{eqnarray} \nonumber
&&M_D=1867 \text{~MeV}~,	\quad~	M_{D_s}=1968 \text{~MeV} ~,\\ \nonumber
&&M_{D^*}=2008 \text{~MeV}~,	\quad  M_{D_s^*}=2112 \text{~MeV}~,\\ \nonumber
&&M_\pi=138 \text{~MeV}~, \qquad  M_K=496\text{~MeV}~,\\
&&M_\eta=548\text{~MeV}~. \label{charmmasses} 
\end{eqnarray}

We will calculate the scattering of the Goldstone bosons off the pseudoscalar
$D$-mesons ($0^+$ channel) as well as off vector $D^*$-mesons ($1^+$ channel).  The
diagrams at the  order we are working are shown in Fig.~\ref{fig:diagrams}.
The diagrams are evaluated in the isospin basis, the relation between the isospin
basis and the  particle basis can be found in the Appendix of Ref.~\cite{Guo:2009ct}.

Evaluating the diagrams for the leading order contact interaction we find that
the contributions for the $0^+$ channel and the $1^+$ channel are the same up to
the different masses for the pseudoscalar and vector charmed mesons, as expected
from heavy quark spin symmetry:
\begin{eqnarray}\label{TLO}
V^{0^+}_{\rm LO}=V^{1^+}_{\rm LO}
=C_0\sqrt{M_1M_3}\frac{1}{2F^2}(E_2+E_4).
\end{eqnarray}
The constants $C_0$ for the various channels are listed in
Tab.~\ref{tab:constants}. Further, $M_1$ and $M_3$ are the masses of the
in-coming and out-going charmed meson, respectively, cf.
Fig.~\ref{fig:diagrams}, and $E_2$ and $E_4$ are the energies of the in-coming and out-going light
mesons.

For the NLO contact interactions  we again find the same contributions except
for the different pseudoscalar and vector charmed meson masses:
\begin{eqnarray}\nonumber
  V_{\rm NLO}^{0^+}&=&V_{\rm NLO}^{1^+}\\
&=&\sqrt{M_1M_3}\frac{1}{2F^2}\bigg[C_1\frac{4}{3}h_1 \nonumber\\
&&+C_{35}(h_3 p_2 \cdot p_4+ 2h_5 E_2 E_4)\bigg] \ ,
\end{eqnarray}
which also preserves spin symmetry. Its S-wave projection is
\begin{eqnarray}\nonumber\label{eq:LO}
 V^{0^+}_{s}&=&V^{1^+}_{s} \\
&=&\sqrt{M_1M_3}\frac{2}{F^2}\bigg[C_0 (E_2+E_4)+C_1\frac{4}{3}h_1  \nonumber\\
&&+C_{35}\frac{1}{M_D}\tilde h_{35}  E_2 E_4\bigg]
\end{eqnarray}
with $h_{35}=h_3+2h_5\equiv {{\tilde h}_{35}/M_D}$. This is the only free
parameter at this order. This is fitted to the mass of the $D_{s0}^*(2317)$. We
get $\tilde h_{35}=0.35$ for the dimensionless parameter, which is of natural
size.

The $s$--channel exchange needs a P-wave interaction and thus does not need to
be considered. The only contribution from exchange diagrams that does not have
an vanishing S-wave projection is the $u$--channel exchange of a charmed
pseudoscalar meson in the $1^+$ channel:
\begin{eqnarray}\label{eq:exch}
V_{\rm PS-Ex}^{1^+}&=&-C_u \frac{4g^2_\pi}{3F^2}
\bigg[\frac{E_4}{M_1}|\vec p|^2+  \frac{E_2}{M_3}|\vec p'|^2\bigg] \non\\
&&\times \frac{1}{(v\cdot k)}\sqrt{M_1M_3},
\end{eqnarray}
where the constant $C_u$ is also given in Tab.~\ref{tab:constants}, and
$\vec{p}$ and $\vec{p}'$ are the three-momenta of the in-coming and out-going
mesons in the center-of-mass frame and the energy transfer $(v\cdot k)$ is defined in Fig.~\ref{fig:diagrams}.
Observe, since $k$, $\vec p$ and $\vec p'$ are counted of order of the $E_i$,
this potential appears formally at NLO.
Since it acts in the $1^+$ channel, while there is no counter--part 
in the $0^+$ channel, it thus provides formally the leading source of spin symmetry violation.
However, in practice its contribution turns out to be very small, due to three
reasons: first of all $g_\pi^2=0.1$, second, $|\vec p|\sim \sqrt{M_K\epsilon}$,
 with $\epsilon\equiv
M_{D^*}+M_K-M_{D_{s1}(2460)}\approx M_{D}+M_K-M_{D_{s0}^*(2317)}$ being the
binding energy, which is significantly smaller than $M_K$ and, most
importantly, it does not operate in the $KD^*\to KD^*$ channel (in that
channel $C_u=0$ --- c.f. Tab.~\ref{tab:constants}). In total it gives
a negligible contribution.

  For SU(3) calculations, the uncertainty
from the chiral expansion up to NLO is $\mathscr
O(M_K^2/\Lambda_\chi^2)$, where $\Lambda_\chi\approx4\pi F_\pi$ sets the hard
scale in the chiral expansion, compared with the LO contributions. So we use
\begin{equation}\label{eq:error}
 V^\pm=\left(1\pm \frac{M_K^2}{\Lambda_\chi^2}\right)V_s,
\end{equation}
where $V_{s}$ is the S-wave projection of the full  NLO amplitude calculated
here for the $0^+$ and $1^+$ channel, respectively. A more detailed discussion
of the scattering amplitudes is given in the Appendix.

\begin{table*}
\centering \caption{Coefficients for the amplitudes for all possible channels
  with total strangeness $S=1$ and total isospin $I=0$. While $C_0$, $C_1$,
  and
$C_{35}$ act in both channels $0^+$ and $1^+$, while $C_u$ acts solely in $1^+$.}
\renewcommand{\arraystretch}{1.3}
\begin{tabular}{|l|l|l|l|l|}
\hline
Channel & $C_0$ & $C_1$ & $C_{35}$ & $C_u$ \\
\hline
 $DK\rightarrow DK$ & $-2$ & $-4M_K^2$ & $\;\;\:2$ & $\;\;\:0$ \\ 
   $D_s\eta\rightarrow D_s\eta$ & $\;\;\:0$ & $-2(2M_\eta^2-M_\pi^2)$ & $\;\;\:\frac 43$  & $\;\;\:\frac 23$ \\ 
   $D_s\eta\rightarrow DK$ & $-\sqrt{3}$ &  $-\frac{\sqrt{3}}{2}(5M_K^2-3M_\pi^2)$ & $\;\;\:\frac{ \sqrt{3}}{ 3}$ &  $-\sqrt{\frac 13}$
 \\\hline
\end{tabular}
\label{tab:constants}
\end{table*}

Dynamical generation of bound states from a theory without bound states as
fundamental fields is a non-perturbative phenomenon. It cannot be provided by
any finite perturbative expansion in the momentum. Thus we have to unitarize, {\it
i.e.} resum, the amplitude we had obtained so far. Following
Ref.~\cite{Oller:2000fj}, the unitarized amplitude can be written as
\begin{equation}\label{Tmatrix}
  T(s)=V^\pm(s)[1-G(s)\cdot V^\pm(s)]^{-1},
\end{equation}
where $G(s)$ is the diagonal matrix with non-vanishing elements being loop
integrals of the relevant channels --- see
Eq.~(\ref{firstloop}) below. The bound state mass is found as a pole of the
analytically continued unitarized scattering matrix $T$. To be specific,
the pole of a bound state, as is the case
here, is located on the real axis below threshold on the first Riemann sheet of the
complex energy plane (note that because we neglect isospin-breaking, these
bound states do not acquire a width). In our approach, both the $D_{s0}^*(2317)$
and the $D_{s1}(2460)$
appear as poles in the $S=1$  isoscalar $I=0$ channel.
Since they couple predominantly to $DK$ and $D^*K$, 
we interpret the $D_{s0}^*(2317)$ and $D_{s1}(2460)$ as  $DK$ and $D^*K$
bound states, respectively.

At this point we have to make some comments on the loop matrix. In previous
works on this subject
\cite{Gamermann:2006nm,Gamermann:2007fi,Guo:2009ct}, 
the relativistic two-particle loop function
\begin{eqnarray}\label{firstloop}
 I{=}i\int \!
 \frac{d^4q}{(2\pi)^4}\frac{1}{q^2{-}M_\phi^2{+}i\epsilon}\frac{1}{(q{-}P)^2{-}M_D^2{+}i\epsilon}
\end{eqnarray}
was used in dimensional regularization with a subtraction constant $a(\mu)$
which absorbs the scale dependence
of the integral, where $M_\phi$ and $M_D$ are the masses of
the light meson and charmed meson involved in the loop and $P$ denotes the
total momentum. However, this relativistic loop function violates spin
symmetry very strongly. Particularly, using the parameters fitted to the mass
of the $D_{s0}^*(2317)$, we find the mass of the $D_{s1}(2460)$ at
2477~MeV. From a field theoretical point of view, using a relativistic
propagator for a heavy meson and using the Lagrangian from heavy quark
expansion simultaneously is inconsistent. There is a well-known problem in
relativistic baryon chiral perturbation theory for pion--nucleon scattering,
which is closely related to the case here. The relativistic nucleon propagator
in a loop explicitly violates power counting, see \cite{Gasser:1987rb}. In our
case this effect appears only logarithmically as $\log\left({M_D^2}/{\mu^2}
\right)$.
However, it induces a violation of spin symmetry in the interaction already at
NLO and thus violates the power counting.

There are different ways to deal with this. In \cite{Kolomeitsev:2003ac} a
subtraction scheme is used that identifies the scale $\mu$ with the mass of the
$D$- and $D^*$-mesons, respectively, and thus the spin symmetry violating terms
disappear. In Refs.~\cite{Gamermann:2006nm,Gamermann:2007fi} simply different
subtraction constants are used for the two channels. In this work we use a
static propagator for the heavy meson, as this is consistent with the
Lagrangian in the
heavy quark expansion. This also allows us to use the
same subtraction constant and scale of regularization for both channels. The
heavy meson propagator for the vector $D^*$-meson then takes the form
\begin{eqnarray}\label{eq:prop}
 \frac{\{P^{\mu\nu},1\}}{q^2-M_D^2+i\epsilon}\rightarrow  \frac{\{P^{\mu\nu},1\}}{2M_D(v\cdot k
   -\Delta+i\epsilon)} \ ,
\end{eqnarray}
where $P^{\mu\nu}$ is a projector for spin-1. Dropping $\Delta$ in the
denominator and using the one in the numerator of Eq.~({\ref{eq:prop}) amounts
to the propagator  for the pseudoscalar $D$-meson.

With this the integral $I$ defined in Eq.~(\ref{firstloop}) turns into
\begin{eqnarray}\nonumber
 I&=&
\frac{1}{16\pi^2
  M_D}\bigg\{(P^0{-}M_D)\left[a(\mu){+}\log\left(\frac{M_\phi^2}{\mu^2}\right)\right]\\  \nonumber
&&{+}2\sqrt{(P^0{-}M_D)^2{-}M_\phi^2}\cosh^{{-}1}\left(\frac{P^0{-}M_D}{M_\phi}\right)\\
&&{-}2\pi
i|\vec p_{\rm cms}| \bigg\}+ \mathcal{O}\left(\frac{M_\phi}{M_D}\right),
\label{nonrelloop}
\end{eqnarray}
with $a(\mu)$ a subtraction constant, $\mu$ the scale of dimensional
regularization, which is fixed to the averaged mass of $D$ and $D^*$, and $|\vec
p_{\rm cms}|=\sqrt{(P_0-M_D)^2-M_\phi^2}$. Note, the use 
of the modified loop function of Eq.~(\ref{nonrelloop}) leads only
to tiny changes from relativistic effects in the observables discussed in
Refs.~\cite{Guo:2008gp,Guo:2009ct}. The subtraction constant $a(\mu)$ is fitted to
the mass of the $D_{s0}^*(2317)$. We find $a(\mu\!= \! 1936~{\rm
MeV})\!=\!-3.034$. Using the same parameters for the axial-vector channel we
find the pole at
\begin{equation}\label{Ds1result}
M_{D_{s1}(2460)}=(2459.6\pm0.6\pm1.8)~{\rm MeV} \ ,
\end{equation}
 where the first uncertainty stems from the experimental
uncertainty for the mass of the $D_{s0}^*(2317)$, which was used for fitting
the parameters, and the second 4\% uncertainty estimates the higher orders
from the heavy quark expansion --- since
there is no significant spin symmetry breaking term in the scattering amplitudes at NLO,
spin symmetry breaking effect should appear at $\mathscr
O([\Lambda_{\rm QCD}/m_c]^2)$. Hence the second uncertainty can be estimated by
$(\Lambda_{\rm QCD}/m_c)^2 \epsilon$. In this context it is interesting to note
that the spin--symmetry violation induced by $M_{D_s*}-M_{D_s}\neq
M_{D*}-M_{D}$, which is of N$^2$LO, indeed contributes to 
$M_{D_{s1}(2460)}$ less than 1 MeV, consistent with the uncertainty estimate.

 The result in Eq.~(\ref{Ds1result}) is in
perfect agreement with the experimental result $$M^{\rm
  exp}_{D_{s1}(2460)}=(2459.5\pm0.6)~{\rm MeV} \ .$$  To the order we are working
instead of $F_\pi$ we could as well have used $F_K = 1.18 F_\pi$. However,
this change does not have any impact on the mass calculated for the
$D_{s1}(2460)$ once the subtraction constant is re-adjusted to reproduce the
mass for the scalar state.

The formalism used here can easily be applied to the meson sector with open
bottom as well. In order to do so, we just have to replace the $D$-mesons with
the corresponding ${\bar B}$-mesons. Their masses are \cite{pdg2010}:
\begin{eqnarray}\nonumber
&& M_{{\bar B}}=5279~{\rm MeV},\quad
M_{{\bar B}_s}=5366~{\rm MeV},\nonumber\\
&& M_{{\bar B}^{*}}=5325~{\rm MeV},\quad
 M_{{\bar B}_s^{*}}=5415~{\rm MeV} \label{bmasses} .
\end{eqnarray}
Heavy quark flavor symmetry implies that all the other parameters stay the
same up to higher order corrections. This allows us to predict the masses of the
${\bar B}K$ and ${\bar B}^*K$ bound states in the ($S=1,I=0$) channel. The predicted
masses are given in Table~\ref{tab:comparison}, together with a comparison with
previous
calculations~\cite{Kolomeitsev:2003ac,Guo:2006fu,Guo:2006rp}.~\footnote{The
$B^{(*)}$ mesons in the calculations of Ref.~\cite{Kolomeitsev:2003ac} should be
understood as ${\bar B}^{(*)}$ mesons which contain a $b$ quark rather than
$\bar b$.} Our results are in reasonable agreement with the previous ones.
\begin{table}[t]
\centering \caption{Comparison of our predictions of the masses of the ${\bar
B}K$ and ${\bar B}^*K$ bound states with those in
Refs.~\cite{Kolomeitsev:2003ac,Guo:2006fu,Guo:2006rp}. All masses are given in
MeV. The uncertainties given in the first column originate from the residual
scale dependence and an estimate of higher order effects, respectively.}
\renewcommand{\arraystretch}{1.3}
\begin{tabular}{|l|l l l|}
\hline
 & This paper & Ref.~\cite{Kolomeitsev:2003ac} & Refs.~\cite{Guo:2006fu,Guo:2006rp} \\
\hline
 $M_{B^*_{s0}}$ & $5696\pm 20 \pm 30$ & $5643$ & $5725\pm39$ \\ 
 $M_{B_{s1}}$   & $5742\pm 20\pm 30$ & $5690$ & $5778\pm7$
 \\ \hline
\end{tabular}
\label{tab:comparison}
\end{table}
 In our calculation we have two sources of uncertainties, both shown
 explicitly in the table. One stems from higher order effects, which may be
 obtained from multiplying the binding energy by 20\%, estimated as ${\cal
   O}(\Lambda_{\rm QCD}/m_c)$ since heavy flavor symmetry was used to relate
 the LECs in charm and bottom sectors. The other one originates from the
 intrinsic scale dependence of the result: To come to a fully renormalization
 group invariant amplitude, a complete one--loop calculation would have been
 necessary for the transition amplitude.  This, however, is beyond the scope
 of this paper. Thus, when connecting the charm to the bottom sector, a
 residual scale dependence remains --- to quantify it we varied the scale
 parameter $\mu$ (see Eq. (\ref{nonrelloop})) from the averaged mass of $D$
 and $D^*$ to that of $\bar B$ and $\bar B^*$, while keeping $a(\mu)$ fixed.
 Combining the two uncertainties in quadrature gives a total uncertainty 40
 MeV which is about 1 \% for the masses.  When again switching from $F_\pi$ to
 $F_K$ as the strength parameter of the interaction (c.f. Eq.~(\ref{Udef}) and
 sentences below), the predicted masses change by 8 MeV only, well consistent
 with our uncertainty estimates.

It should be stressed that the uncertainties quoted for $M_{B^*_{s0}}$ and
$M_{B_{s1}}$ are highly correlated. As already explained for the charmed
system, within the molecular scenario the relation
\begin{equation}
M_{B_{s1}}{-}M_{B_{s0}^*}\simeq M_{B^*}{-}M_B \ 
\end{equation}
should hold up to corrections of $\mathscr O([\Lambda_{QCD}/m_b]^2)$ --- c.f.
discussion below Eqs.~(\ref{eq:exch}) and (\ref{Ds1result}). Thus we predict
\begin{equation}\label{split}
M_{B_{s1}}{-}M_{B_{s0}^*}=46\pm 0.4 \pm 1 \ \mbox{MeV} \ , 
\end{equation}
where the first uncertainty comes from the current experimental uncertainty in
$M_{B^*}{-}M_B$ and the second from the estimated spin breaking effects in the
formation of the molecule. Clearly, all mass differences deduced from
Tab.~\ref{tab:comparison} are consistent with this value.

$M_{B_{s1}}$ and $M_{B_{s0}^*}$ have not been measured so far. Note that
their existence in the deduced mass range and, especially, with the mass
splitting of Eq.~(\ref{split}), is a crucial and highly non--trivial test for
the theory presented and especially for the molecular nature of both states.

\section{Chiral extrapolations}\label{sec:chextra}

To test the nature of the resonances besides further experimental data we can
also compare our results to lattice calculations (for a corresponding study
in the light meson sector see Ref.~\cite{qmdepsigmarho}). To do so, we extend
the calculations from the physical world to unphysical quark masses which are
frequently used in lattice calculations.
Varying the light quark masses is equivalent to varying the pion mass.
Although the physical strange quark mass is nowadays routinely used in lattice
calculations, we emphasize that by varying the strange quark mass, or
equivalently varying the kaon mass, one can learn a lot about the nature of
some hadrons, as will be discussed below. So we have to express all results in
terms of pion and kaon masses, respectively.
In the following chapter only the charmed mesons are discussed explicitly,
however, analogous arguments apply to their bottom partners.

In order to proceed we need to assume that the subtraction constant $a(\mu)$
does not depend on the light quark masses. We stress, however, that even
allowing for a quark mass dependence of $a(\mu)$ would not change the general
features of the results, but might slightly enhance the uncertainties.

\subsection{Pion mass dependence}

\begin{figure*}
\begin{minipage}{0.47\linewidth}
\centering
\includegraphics[width=1\linewidth]{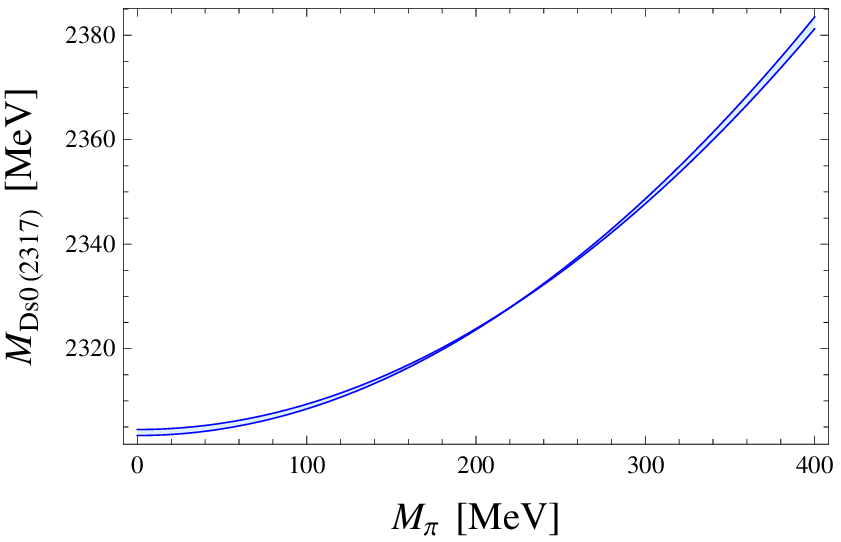}
\end{minipage}
\hspace{0.5cm}
\begin{minipage}{0.47\linewidth}
\centering
\includegraphics[width=1\linewidth]{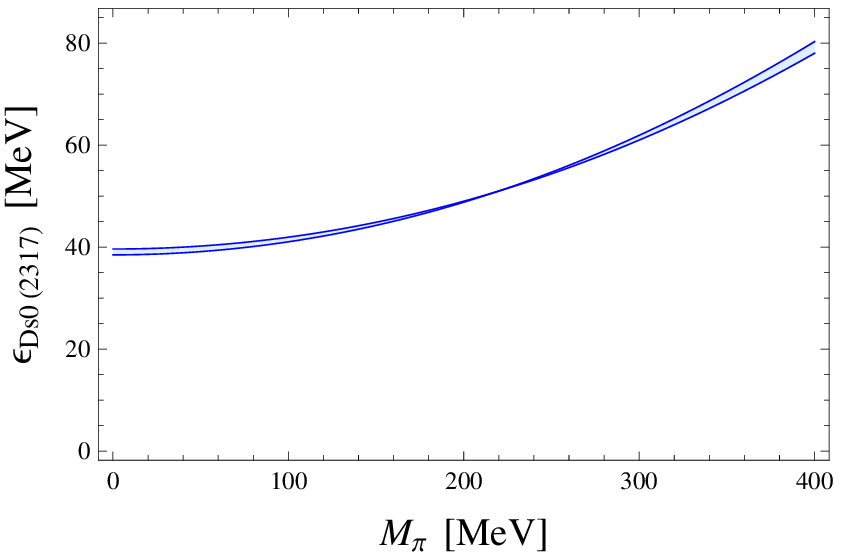}
\end{minipage}

\vspace{1cm}

\begin{minipage}{0.47\linewidth}
\centering
\includegraphics[width=1\linewidth]{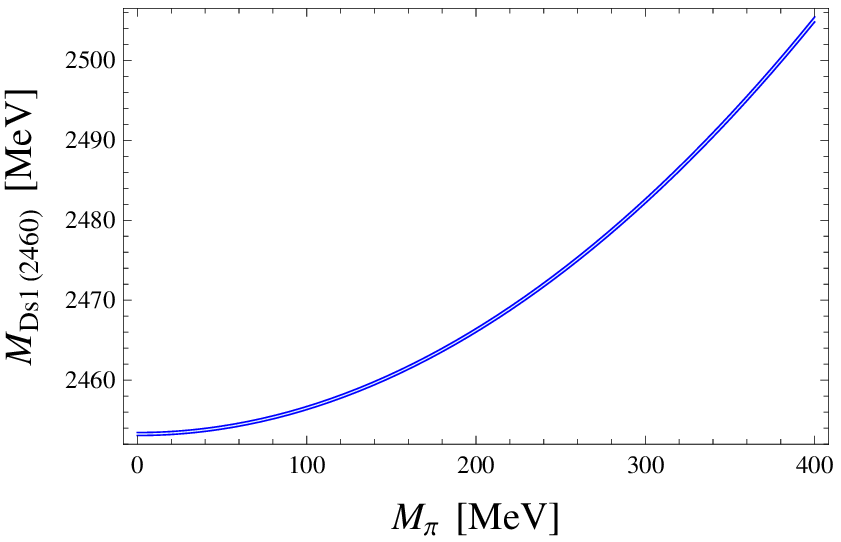}
\end{minipage}
\hspace{0.5cm}
\begin{minipage}{0.47\linewidth}
\centering
\includegraphics[width=1\linewidth]{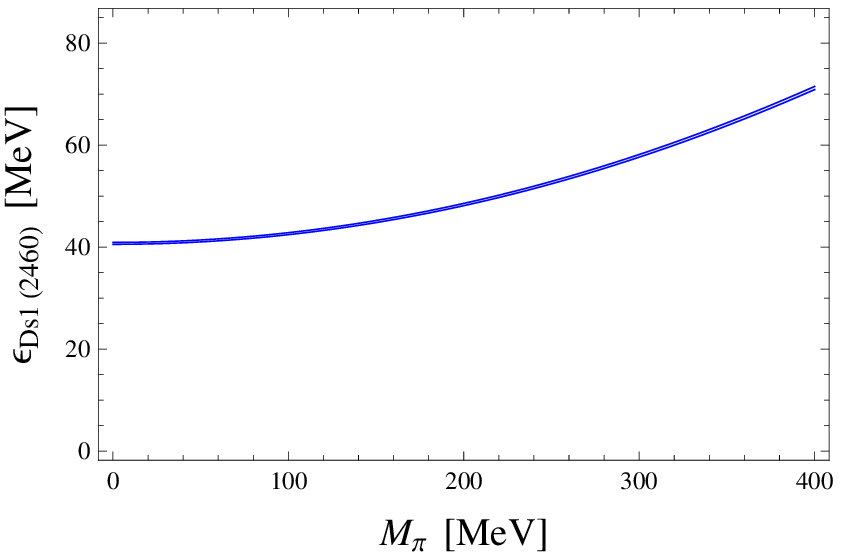}
\end{minipage}
\caption{The masses and binding energies of the $D^*_{s0}(2317)$ and the $D_{s1}(2460)$
as a function of the pion mass.} \label{fig:PlotsPion}
\end{figure*}


Lattice QCD calculations are often performed at larger quark masses than
realized in nature --- for the $KD$ system of interest here exploratory
lattice
studies are presented in Ref.~\cite{Liu:2008rz}.
In addition, as we will argue here, the quark mass dependence of a state
contains important information on its nature. Varying $u$ and $d$
quark masses can be expressed by varying the pion mass. In \cite{Jenkins:1992hx},
the charmed meson masses were expanded up to one loop order in the chiral expansion.
Nevertheless for our purposes the expansion up to $\mathscr O(M_\pi^2)$ is
sufficient.

Using the Lagrangian given in Eq.~\eqref{Eq:LagrNLO} we find the NLO correction
to the charmed meson masses to be
\begin{eqnarray}
 \delta M_{D^{(*)}}^2=4 h_1 B \hat m \ , \qquad \delta M_{D^{(*)}_s}^2=4 h_1 B
 m_s \ ,
\end{eqnarray}
with $\hat m = (m_u+m_d)/2$ the average light quark mass.
When studying the pion mass dependence, we consider the physical value for the
strange quark mass here, thus we can use the physical mass for the
$D_s^{(*)}$. Using $M_\pi^2=2B\hat m$ yields, up to $\mathscr
O(M_\pi^2)$,
\begin{eqnarray}
M_{D^{(*)}} = \overset{_\circ}{M}_{D^{(*)}} + h_1\frac{M_\pi^2}{\overset{_\circ}{M}_{D^{(*)}}},
\end{eqnarray}
where $\overset{_\circ}{M}_{D^{(*)}}$ is the charmed meson mass in the SU(2)
chiral limit $m_u=m_d=0$ with $m_s$ fixed at its physical value. For the kaon
and the eta mass we can find similar expressions by using ${M_K^2=B(\hat
m+m_s)}$, ${\overset{_\circ}{M}_K^2=Bm_s}$ and ${M_\eta^2=B\frac{2}{3}(\hat
m+2m_s)}$, ${\overset{_\circ}{M}_\eta^2=B\frac{4}{3}m_s}$ from the LO chiral
expansion:
\begin{eqnarray}
M_K = \overset{_\circ}{M}_K + \frac{M_\pi^2}{4\overset{_\circ}{M}_K}, \qquad
M_\eta = \overset{_\circ}{M}_\eta + \frac{M_\pi^2}{6\overset{_\circ}{M}_\eta}.
\end{eqnarray}

In Fig.~\ref{fig:PlotsPion} we show the mass of both resonances,
$D_{s0}^*(2317)$ and $D_{s1}(2460)$, as well as their binding energies as a
function of the pion mass. Note that the observed rather strong pion mass
dependence is specific for a molecular state. The corresponding dependence for a
quark state should be a lot weaker. To understand this, one notices that a pure
$c\bar s$ state does not contain any $u,~\bar u,~d,~\bar d$ quarks. 
The leading
term containing these light quarks is $1/N_c$ suppressed.  Thus, for the
quark state, the pion mass dependence can only enter via $D^{(*)}K$ loops ---
as in case of the molecular state. However, their contribution should be a lot
smaller for the quark state than for the molecular state. To see this, observe
that the loop contributions from a particular meson pair  is
proportional to the squared coupling of that meson pair to the re\-so\-nance. As
shown in Refs.~\cite{Weinberg:1965zz,Baru:2003qq}, this coupling is
proportional to the probability to find the molecular state in the physical
state.  Thus, the pion mass dependence for $D_{s0}^*(2317)$ and $D_{s1}(2460)$
should be maximal if both are pure molecules. This case is depicted
in Fig.~\ref{fig:PlotsPion}. On the other hand it should be very weak for the
admittedly unrealistic scenario of a pure quark state.  Note that the
mentioned relation between the coupling and the structure of the state holds
only in leading order in an expansion in $\sqrt{M_K\epsilon}/\beta$ (see
Ref.~\cite{molgamgam} for a detailed discussion), with the inverse range of
forces $\beta\sim m_\rho$. For the scalar and axial
vector open charm mesons this gives an uncertainty of the order of 20\%.
As a result of the larger binding, for the bottom analogs we even find 30\%.
In addition in the analysis we neglected terms of ${\cal{O}}(1/N_c)$. 
Thus, from this kind of analysis at most statements like `the state is
predominantly molecular/compact' are possible.

Furthermore we notice that the plots show an almost identical behavior in the
scalar and the axial-vector channel. So the spin symmetry breaking effects are
only very weak here.  We see that the binding energy in both cases varies from
about 40 MeV to about 80 MeV.


\subsection{Kaon mass dependence}

Before going into details of calculations, let us make some general
statements about  the $M_K$--dependence of the mass of a bound state
of a kaon and some other hadron. The mass of such a kaonic bound state
is given by
\begin{equation}
M = M_K+M_h-\epsilon,
\end{equation}
where $M_h$ is the mass of the other hadron, and $\epsilon$ denotes the binding
energy. Although both $M_h$
and $\epsilon$ have some kaon mass dependence, it is expected to be a lot
weaker
than that of the kaon itself.
Thus, the important implication of this simple formula is that the leading
kaon mass dependence of a kaon--hadron bound state is {\em linear, and the slope is unity}. The only
exception to this argument is if the other hadron
is also a kaon or anti-kaon.~\footnote{The $f_0(980)$ was proposed to be such a
$K\bar K$ bound state~\cite{Weinstein:1990gu,Baru:2003qq}.} In this case, the
leading kaon mass dependence is still linear but with the slope being changed to two.
Hence, as for the $DK$ and $D^*K$ bound states discussed here, we expect that
their masses are linear in the kaon mass, and the slope is approximately one. As we
will see, our explicit calculations confirm this expectation.

\begin{figure*}

\begin{minipage}{0.45\linewidth}
\centering
\includegraphics[width=1\linewidth]{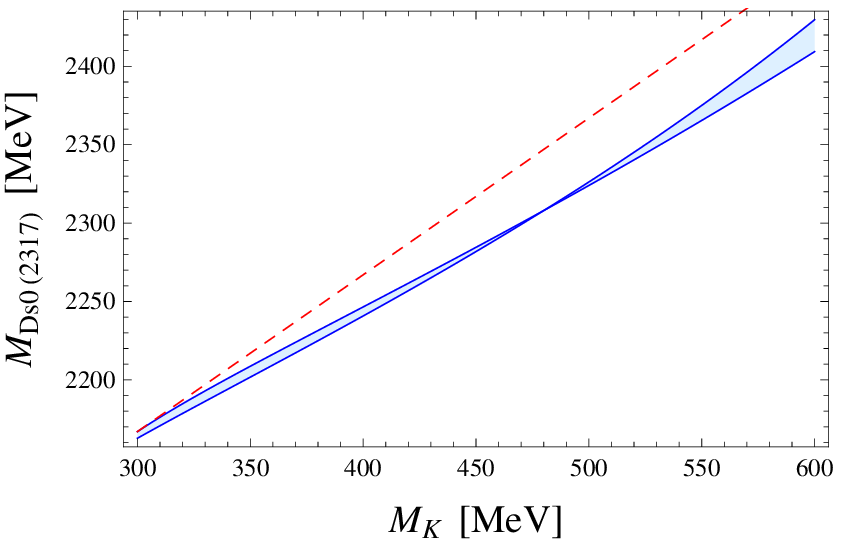}
\end{minipage}
\hspace{1cm}
\begin{minipage}{0.45\linewidth}
\centering
\includegraphics[width=1\linewidth]{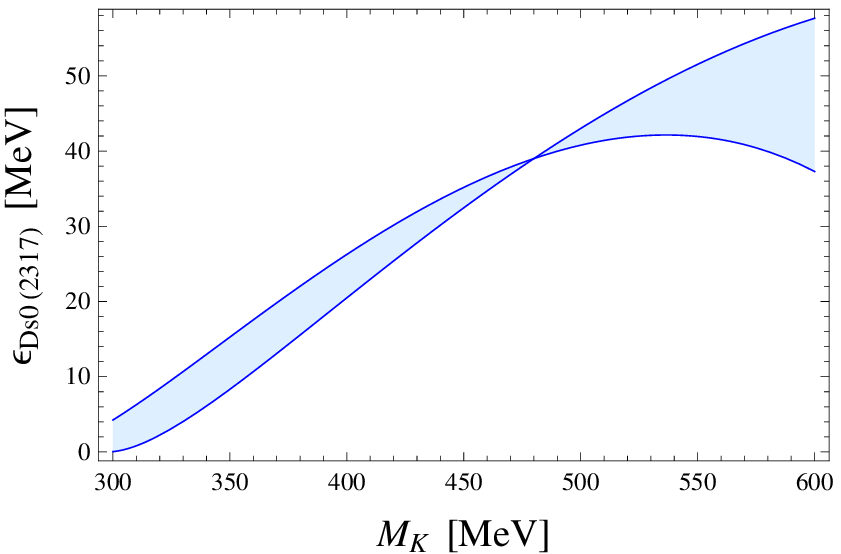}
\end{minipage}

\vspace{1cm}

\begin{minipage}{0.45\linewidth}
\centering
\includegraphics[width=1\linewidth]{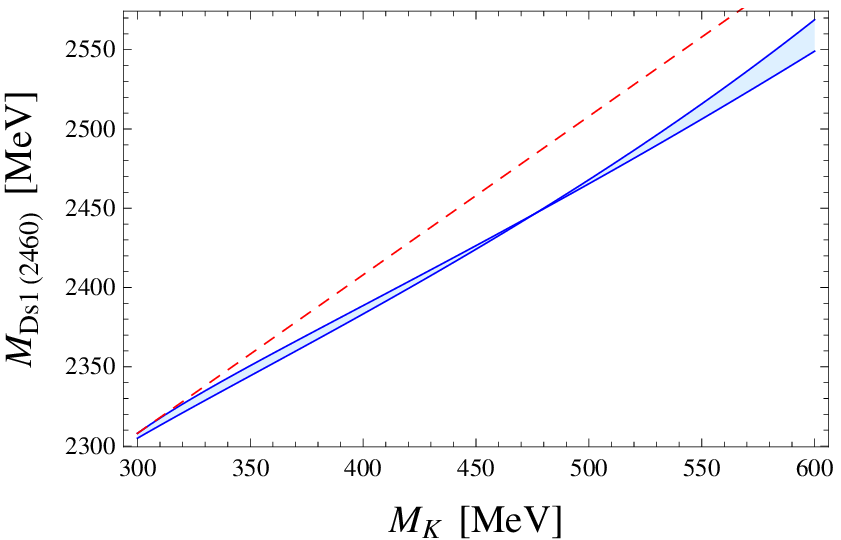}
\end{minipage}
\hspace{1cm}
\begin{minipage}{0.45\linewidth}
\centering
\includegraphics[width=1\linewidth]{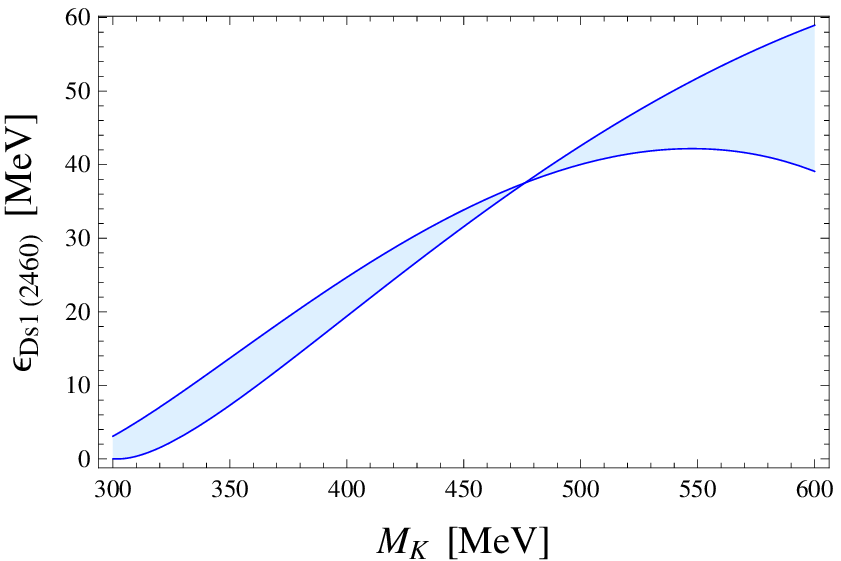}
\end{minipage}
\caption{The blue band shows the masses and binding energies of
  $D^*_{s0}(2317)$ and $D_{s1}(2460)$ in dependence of the kaon mass, the red
  dashed line shows the threshold.}
\label{fig:PlotsKaon}
\end{figure*}

For calculating the kaon mass dependence, we use physical masses for the pion
and the charmed mesons without strangeness.~\footnote{Certainly, lattice
simulations use unphysical values for up and down quarks. However, the
conclusion of this subsection will not be affected by the value of pion mass.
For definiteness, we choose to use physical masses for the pion and non-strange
charmed mesons.} To express the strange quark mass dependence in terms
of kaon masses, we write
\begin{eqnarray}
 M_K^2=B\left(m_s+\hat m\right)~, \quad
\overset{_\circ}{M}_K^2=B\hat m= \frac{1}{2}M_\pi^2~,
\end{eqnarray}
where $\overset{_\circ}{M}_K^2$ is the mass of the kaon in the limit $m_s=0$.
The charmed strange meson mass is
\begin{eqnarray}
M_{D^{(*)}_s} = \overset{_\circ}{M}_{D_s^{(*)}} + 2 h_1\frac{M_K^2}{\overset{_\circ}{M}_{D_s^{(*)}}}-h_1\frac{M_\pi^2}{\overset{_\circ}{M}_{D_s^{(*)}}},
\label{mqdepofD}
\end{eqnarray}
where $\overset{_\circ}{M}_{D_s^{(*)}}$ is the charmed strange meson mass in the
limit $m_s=0$. Finally the eta mass is given to this order by the
Gell-Mann--Okubo relation
\begin{eqnarray}
 M_\eta^2=B\left(\frac 43 m_s+\hat m\right)=\frac 43 M_K^2-\frac 13 M_\pi^2.
\end{eqnarray}

For analyzing the kaon mass dependence we calculate the masses of the $D_{s0}^*(2317)$ and
$D_{s1}(2460)$ and their binding energies as before. The corresponding results
are shown in Fig.~\ref{fig:PlotsKaon}. Furthermore we also show the $D^{(*)}K$
threshold in dependence on the kaon mass.

For the $D_{s0}^*(2317)$, we find that the kaon mass dependence of the mass is
almost perfectly linear, especially in the region $M_K=300 - 500~{\rm MeV}$. For
higher strange quark masses the chiral expansion  is no longer valid and the
results become less reliable.
The kaon mass dependence of the mass and the $DK$ threshold have a similar
slope. The slope of the threshold is one, since the $D$-meson mass is
independent of $M_K$ to the order we are working, and the slope of 
$M_{D^*_{s0}(2317)}$ is about 0.85. This finding is perfectly consistent with
the expectation formulated in the beginning of this subsection. The deviation
of the slope from one can be understood from the kaon mass dependence of the
binding energy (see right column of Fig.~\ref{fig:PlotsKaon}) and from the
effects of the coupling to the $D_s\eta$ channel. The situation for the
$D_{s1}(2460)$ is similar.

On the contrary, if $D_{s0}^*(2317)$ and $D_{s1}(2460)$ are assumed to be quark
states, their masses necessarily depend quadratically on the kaon mass ---
analogous to Eq.~(\ref{mqdepofD}). Thus, an extraction of the kaon mass
dependence of the masses of $D_{s0}^*(2317)$ and $D_{s1}(2460)$ from lattice
data is of high interest to pin down the nature of these states.

\section{Summary}\label{sec:summary}

In this paper we used unitarized heavy meson chiral perturbation theory to
calculate the masses of the resonances $D_{s0}^*(2317)$ and $D_{s1}(2460)$
as well as their bottom analogs $B_{s0}^*$ and $B_{s1}$.
The results for the latter turned out to be consistent with previous analyses.
Especially we predict $M_{B_{s1}}{-}M_{B_{s0}^*}=46\pm 1 \ \mbox{MeV}
\ ,$ where we added the uncertainties of Eq.~(\ref{split}) in quadrature.
In addition,
 calculations were performed with varying pion and  kaon masses. Within the
molecular picture used here, we found a linear dependence of the resonance masses on
the kaon mass with the slope being approximately unity and a strong quadratic dependence
on the pion mass. We argued that, if both states are assumed to be elementary
quark states, the mentioned dependences would be qualitatively different: while
the pion mass dependence should still be quadratic, it should be  a lot weaker,
the kaon mass dependence should come out quadratic as well. Thus, especially an
investigation of the kaon mass dependence of the  $D_{s0}^*(2317)$ and
$D_{s1}(2460)$ or their bottom analogs  within lattice QCD would be very interesting in order to
determine the nature of these resonances.

\medskip

\section*{Acknowledgments}
This work is partially supported by  the Helm\-holtz Association through funds
provided to the Virtual Institute ``Spin and strong QCD'' (VH-VI-231) and by the
DFG (TR 16, ``Subnuclear Structure of Matter''), the European
Community-Research Infrastructure Integrating Activity ``Study of Strongly
Interacting Matter'' (acronym HadronPhysics2, Grant  Ag\-ree\-ment n. 227431) under
the FP7 of the EU, and the BMBF (grant 06BN9006).

\medskip

\begin{appendix}

\section{Amplitudes}

Here we give all the contributing amplitudes and their S-wave projections. In
all the following expressions, $p_{1(3)}$ and $p_{2(4)}$ denote the momentum of
the in-coming (out-going) heavy meson and light meson, respectively, the heavy
meson mass is given by $M_{1(3)}$, and the energy of the light meson is given by
$E_{2(4)}$. The leading order contact interactions are purely S-wave by
construction and identical up to the masses for both channels:
\begin{eqnarray}
  T^{0^+}_{\rm LO}&=&\sqrt{M_1M_3}C_0\frac{1}{2F^2}(E_2+E_4)
\end{eqnarray}
\begin{eqnarray}\nonumber
T^{1^+}_{\rm LO}&=&-\sqrt{M_1M_3}(\epsilon\cdot\epsilon^*) C_0\frac{1}{2F^2}(E_2+E_4)\\
&=&\sqrt{M_1M_3}C_0\frac{1}{2F^2}(E_2+E_4),
\end{eqnarray}
where $\epsilon$ is the polarization vector of the $D^*$.

The NLO contact interaction reads:
\begin{eqnarray}\nonumber
  T_{\rm NLO}^{0^+}&=&\sqrt{M_1M_3}\bigg[C_1\frac{2}{3F^2}h_1\\ 
&&+C_{35}\frac{2}{F^2}(h_3 p_2 \cdot p_4+ 2h_5 E_2 E_4)\bigg]
\end{eqnarray}
\begin{eqnarray}
\nonumber
T_{\rm NLO}^{1^+}&=&-\sqrt{M_1M_3}\epsilon\cdot\epsilon^*\bigg[C_1\frac{2}{3F^2}h_1\\ \nonumber
&&+C_{35}\frac{2}{F^2}(h_3 p_2 \cdot p_4+ 2h_5 E_2 E_4)\bigg]\\ \nonumber
&=&\sqrt{M_1M_3}\bigg[C_1\frac{2}{3F^2}h_1\\
&&+C_{35}\frac{2}{F^2}(h_3 p_2 \cdot p_4+ 2h_5 E_2 E_4)\bigg].
\end{eqnarray}

The matrix element for the exchange of charmed vector mesons in the $1^+$
channel is zero at this order:
\begin{eqnarray} \nonumber
  T^{1^+}_{\rm Vector-Ex}&=&\frac{4g_\pi^2}{F^2}\frac{1}{2v\cdot k}(\vec \epsilon^*\cdot \vec
p\vec \epsilon\cdot \vec p'-\vec\epsilon^*\cdot \vec\epsilon \vec
p\cdot \vec p')\\
&=&0,
\end{eqnarray}
where $\vec \epsilon$ is the spatial component of the polarization vector of the
$D^*$, and $\vec{p}^{(\prime)}$ denotes the three-momentum of the in-coming
(out-going) heavy meson in the center-of-mass frame and $k$ is the residual
momentum of the exchanged heavy meson (c.f. Fig.~\ref{fig:diagrams}).

The exchange of charmed vector mesons in the $0^+$ channel gives a non-vanishing
contribution:
\begin{equation}
   T_{\rm Vector-Ex}^{0^+} = -8C_u\sqrt{M_1M_3} \frac{g^2_\pi}{F^2} \frac{\vec p\cdot \vec p'}{2v\cdot k}
\end{equation}
But since it is purely P-wave, the S-wave projection vanishes. Finally we
consider the exchange of charmed pseudoscalar mesons in the $1^+$ channel:
\begin{eqnarray}\nonumber
T_{\rm PS-Ex}^{1^+}&=&-C_u\frac{8g^2_\pi}{F^2}(\epsilon^*\cdot p_2)(\epsilon\cdot
p_4)\frac{1}{2v\cdot k}\sqrt{M_1M_3}\\ \nonumber
&=&-C_u \frac{8g^2_\pi}{F^2}\bigg[\frac{E_4}{M_1}|\vec p|^2 \cos^2\theta+\frac{E_2}{M_3}|\vec p'|^2\cos^2\theta'\\ \non
&&+\left(\frac{E_2E_4}{M_1M_3}|\vec p||\vec p'|+|\vec p||\vec
p'|\right)\cos\theta\cos\theta' \bigg]\\
&&\times \frac{1}{2(v\cdot k)}\sqrt{M_1M_3},
\end{eqnarray}
where $\theta~ (\theta')$ is the angle between $\vec p~(\vec p')$ and the
quantization axis used for the polarization vectors. The projections $\lambda$
and $\lambda'$ of the polarization vectors is conserved for $S$-wave
scattering. The expression given holds for $\lambda=\lambda'=0$. This gives
the only non-vanishing S-wave projection from the exchange diagrams:
\begin{eqnarray}
T_{\rm PS-Ex}^{1^+}&=&-C_u \frac{4g^2_\pi}{3F^2}\bigg(\frac{E_4}{M_1}|\vec p|^2+  \frac{E_2}{M_3}|\vec p'|^2\bigg) \non\\
&&\times \frac{1}{(v\cdot k)}\sqrt{M_1M_3} \ .
\end{eqnarray}
As explained in the main text, this contribution is formally of NLO, however,
for the resonances discussed it gives a negligible contribution. 
\end{appendix}


\begin{thebibliography}{99}

\bibitem{Aubert:2003fg}
  B.~Aubert {\it et al.}  [BABAR Collaboration],
  Phys.\ Rev.\ Lett.\  {\bf 90}, 242001 (2003)
  (arXiv:hep-ex/0304021).

\bibitem{Besson:2003cp}
  D.~Besson {\it et al.}  [CLEO Collaboration],
  Phys.\ Rev.\  D {\bf 68}, 032002 (2003)
  [Erratum-ibid.\  D {\bf 75}, 119908 (2007)]
  (arXiv:hep-ex/0305100).

\bibitem{Godfrey:1985xj}
  S.~Godfrey and N.~Isgur,
  Phys.\ Rev.\  D {\bf 32}, 189 (1985).

\bibitem{Bardeen:2003kt}
  W.~A.~Bardeen, E.~J.~Eichten and C.~T.~Hill,
  Phys.\ Rev.\  D {\bf 68}, 054024 (2003)
  (arXiv:hep-ph/0305049).

\bibitem{Nowak:2003ra}
  M.~A.~Nowak, M.~Rho and I.~Zahed,
  Acta Phys.\ Polon.\  B {\bf 35}, 2377 (2004)
  (arXiv:hep-ph/0307102).

\bibitem{Mehen:2005hc}
  T.~Mehen and R.~P.~Springer,
  Phys.\ Rev.\  D {\bf 72}, 034006 (2005)
  (arXiv:hep-ph/0503134).

\bibitem{Guo:2009id}
  F.-K.~Guo, C.~Hanhart and U.-G.~Mei{\ss}ner,
  Phys.\ Rev.\ Lett.\  {\bf 102}, 242004 (2009)
  (arXiv:0904.3338 [hep-ph]).

\bibitem{Barnes:2003dj}
  T.~Barnes, F.~E.~Close and H.~J.~Lipkin,
  Phys.\ Rev.\  D {\bf 68}, 054006 (2003)
  (arXiv:hep-ph/0305025).

\bibitem{vanBeveren:2003kd}
  E.~van Beveren and G.~Rupp,
  Phys.\ Rev.\ Lett.\  {\bf 91}, 012003 (2003)
  (arXiv:hep-ph/0305035).

\bibitem{Swanson:2006st}
  E.~S.~Swanson,
  Phys.\ Rept.\  {\bf 429}, 243 (2006)
  (arXiv:hep-ph/0601110).

\bibitem{Zhu:2007wz}
  S.~L.~Zhu,
  Int.\ J.\ Mod.\ Phys.\  E {\bf 17}, 283 (2008)
  (arXiv:hep-ph/0703225).

\bibitem{Kolomeitsev:2003ac}
  E.~E.~Kolomeitsev and M.~F.~M.~Lutz,
  Phys.\ Lett.\  B {\bf 582}, 39 (2004)
  (arXiv:hep-ph/0307133).

\bibitem{Hofmann:2003je}
  J.~Hofmann and M.~F.~M.~Lutz,
  Nucl.\ Phys.\  A {\bf 733}, 142 (2004)
  (arXiv:hep-ph/0308263).

\bibitem{Guo:2006fu}
  F.-K.~Guo, P.~N.~Shen, H.~C.~Chiang, R.~G.~Ping and B.~S.~Zou,
  Phys.\ Lett.\  B {\bf 641}, 278 (2006)
  (arXiv:hep-ph/0603072).

\bibitem{Guo:2006rp}
  F.-K.~Guo, P.~N.~Shen and H.~C.~Chiang,
  Phys.\ Lett.\  B {\bf 647}, 133 (2007)
  (arXiv:hep-ph/0610008).

\bibitem{Gamermann:2006nm}
  D.~Gamermann, E.~Oset, D.~Strottman and M.~J.~Vicente Vacas,
  Phys.\ Rev.\  D {\bf 76}, 074016 (2007)
  (arXiv:hep-ph/0612179).

\bibitem{Gamermann:2007fi}
  D.~Gamermann and E.~Oset,
  Eur.\ Phys.\ J.\  A {\bf 33}, 119 (2007)
  (arXiv:0704.2314 [hep-ph]).

\bibitem{Faessler:2007gv}
  A.~Faessler, T.~Gutsche, V.~E.~Lyubovitskij and Y.~L.~Ma,
  Phys.\ Rev.\  D {\bf 76}, 014005 (2007)
  (arXiv:0705.0254 [hep-ph]).

\bibitem{Lutz:2007sk}
  M.~F.~M.~Lutz and M.~Soyeur,
  Nucl.\ Phys.\  A {\bf 813}, 14 (2008)
  (arXiv:0710.1545 [hep-ph]).

\bibitem{Guo:2008gp}
  F.-K.~Guo, C.~Hanhart, S.~Krewald and U.-G.~Mei{\ss}ner,
  Phys.\ Lett.\  B {\bf 666}, 251 (2008)
  (arXiv:0806.3374 [hep-ph]).

\bibitem{Colangelo:2003vg}
  P.~Colangelo and F.~De Fazio,
  Phys.\ Lett.\  B {\bf 570}, 180 (2003)
  (arXiv:hep-ph/0305140).

\bibitem{Guo:2009ct}
  F.-K.~Guo, C.~Hanhart and U.-G.~Mei{\ss}ner,
  Eur.\ Phys.\ J.\  A {\bf 40}, 171 (2009)
  (arXiv:0901.1597 [hep-ph]).


\bibitem{Weinberg:1965zz}
  S.~Weinberg,
  Phys.\ Rev.\  {\bf 137}, B672 (1965).

\bibitem{Baru:2003qq}
  V.~Baru, J.~Haidenbauer, C.~Hanhart, Yu.~Kalashnikova and A.~E.~Kudryavtsev,
  Phys.\ Lett.\  B {\bf 586}, 53 (2004)
  (arXiv:hep-ph/0308129).

\bibitem{Burdman:1992gh}
  G.~Burdman and J.~F.~Donoghue,
  Phys.\ Lett.\  B {\bf 280}, 287 (1992).

\bibitem{Wise:1992hn}
  M.~B.~Wise,
  Phys.\ Rev.\  D {\bf 45}, 2188 (1992).

\bibitem{Yan:1992gz}
  T.~M.~Yan, H.~Y.~Cheng, C.~Y.~Cheung, G.~L.~Lin, Y.~C.~Lin and H.~L.~Yu,
  Phys.\ Rev.\  D {\bf 46}, 1148 (1992);
  {\bf 55}, 5851 (1997)(E).

\bibitem{Falk:1990yz}
  A.~F.~Falk, H.~Georgi, B.~Grinstein and M.~B.~Wise,
  Nucl.\ Phys.\  B {\bf 343}, 1 (1990).

\bibitem{pdg2010}
 K.~Nakamura {\it et al.} [Particle Data Group],
 J.\ Phys.\ G {\bf 37}, 075021 (2010).

\bibitem{Oller:2000fj}
  J.~A.~Oller and U.-G.~Mei{\ss}ner,
  Phys.\ Lett.\  B {\bf 500}, 263 (2001)
  (arXiv:hep-ph/0011146).

\bibitem{Gasser:1987rb}
  J.~Gasser, M.~E.~Sainio and A.~Svarc,
  Nucl.\ Phys.\  B {\bf 307}, 779 (1988).


\bibitem{qmdepsigmarho} 
 C.~Hanhart, J.~R.~Pelaez and G.~Rios,
  Phys.\ Rev.\ Lett.\  {\bf 100} (2008) 152001
  [arXiv:0801.2871 [hep-ph]].
 

\bibitem{Liu:2008rz}
  L.~Liu, H.~W.~Lin and K.~Orginos,
  PoS {\bf LATTICE2008}, 112 (2008)
  (arXiv:0810.5412 [hep-lat]).

\bibitem{Jenkins:1992hx}
  E.~E.~Jenkins,
  Nucl.\ Phys.\  B {\bf 412}, 181 (1994)
  (arXiv:hep-ph/9212295).

\bibitem{molgamgam}
  C.~Hanhart, Yu.~S.~Kalashnikova, A.~E.~Kudryavtsev and A.~V.~Nefediev,
  Phys.\ Rev.\  D {\bf 75} (2007) 074015
  [arXiv:hep-ph/0701214].


\bibitem{Weinstein:1990gu}
  J.~D.~Weinstein and N.~Isgur,
  Phys.\ Rev.\  D {\bf 41}, 2236 (1990).

\end{thebibliography}

\end{document}